\documentclass[prd,showpacs,twocolumn,preprintnumbers,amsmath,amssymb,superscriptaddress,floatfix,nofootinbib]{revtex4-2}%

\usepackage{graphicx}
\usepackage{bm}
\usepackage{amsmath}
\usepackage{amsfonts}
\usepackage{amssymb}
\usepackage{color}
\usepackage{multirow}
\usepackage{subfigure}
\usepackage{epstopdf}
\usepackage[colorlinks, citecolor=blue,anchorcolor=red,menucolor=red, linkcolor=red,filecolor=red,runcolor=red,urlcolor=blue,frenchlinks=red]{hyperref}

\begin{document}
	\title{Searching for a $P_{cs}(4200)$ state in the $\Lambda_b\to\phi\eta_c\Lambda$ reaction}

	\author{Wen-Tao Lyu}\email{lvwentao9712@163.com}
	\affiliation{School of Physics, Zhengzhou University, Zhengzhou 450001, China}
	\affiliation{Departamento de Física Teórica and IFIC, Centro Mixto Universidad de Valencia-CSIC Institutos de Investigación de Paterna, 46071 Valencia, Spain}
	\vspace{0.5cm}
	
	\author{Eulogio Oset}\email{oset@ific.uv.es}
	\affiliation{Departamento de Física Teórica and IFIC, Centro Mixto Universidad de Valencia-CSIC Institutos de Investigación de Paterna, 46071 Valencia, Spain}
	\vspace{0.5cm}
	
	

\begin{abstract}
	
We propose the $\Lambda_b\to\phi \eta_c \Lambda$ reaction to observe a $P_{cs}$ state around $4200$~MeV, predicted at lower masses than expected from comparison with the $P_c$ states, stemming as a consequence of the important role played by coupled channels in the $P_{cs}$ case, which does not appear in the $P_c$ case. That state decays to $\eta_c \Lambda$ with a width of about $200$~keV. The reaction is related to  $\Lambda_b^0\to\phi D_s^-  \Lambda_c^+$, which has already been observed. We predict a branching fraction for  $\Lambda_b\to\phi P_{cs}(4200)$; $P_{cs}\to\eta_c \Lambda$ of the order of $10^{-5}$, which is within present capabilities of the LHCb collaboration. The observation of this state would bring valuable light on the nature of the $P_c$ and $P_{cs}$ states and the role played by coupled channels in hadron structure and hadron reactions. 

\end{abstract}
	
	\date{\today}
	
	\maketitle
	
\section{Introduction}\label{sec1}

The discovery of the pentaquark $P_c$ and $P_{cs}$ states gave a boost to hadron physics, showing evidence of baryon states with more than three quarks, instead of three in conventional baryons~\cite{LHCb:2015yax,LHCb:2019kea,LHCb:2020jpq,LHCb:2022ogu,LHCb:2021chn}. The discovery of these states gave a boost to the molecular picture of many baryon states, given the accurate prediction of the existence of these states based on the interaction of mesons and baryons~\cite{Wu:2010jy,Wu:2010vk,Wu:2012md,Xiao:2013yca,Karliner:2015ina}. There has been extensive work on the issue after the discovery of these states, which is reviewed in different papers~\cite{Chen:2016qju,Lebed:2016hpi,Guo:2017jvc,Liu:2019zoy,Ali:2017jda,Chen:2022asf}.

In the present work we concentrate on the $P_{cs}$ states where there is still some debate concerning their nature. While there is a broad consensus about the $P_c$ states, this is not the case of the $P_{cs}$ states, as we shall see. The $P_c(4312)$ state is associated to a $\bar{D}\Sigma_c$ molecule, while $P_c(4440)$, $P_c(4457)$ are associated to the $\bar{D}^*\Sigma_c$ molecule with $J^P=1/2^-$, $3/2^-$, with a small mixture with other coupled channels, and there is still some debate about the spin $1/2^-$ or $3/2^-$ of the $P_c(4440)$ and $P_c(4457)$ states~\cite{Yang:2024nss,Xiao:2019aya,Chen:2021cfl,Lin:2023ihj,Yalikun:2021bfm,Liu:2020hcv,Peng:2020hql,Xiao:2020frg,Peng:2020gwk,Xu:2020gjl,Du:2021bgb,PavonValderrama:2019nbk,Yamaguchi:2019seo,Liu:2019zvb,Meng:2019ilv,Xiao:2019aya,He:2019ify,Chen:2019asm,Du:2021fmf,Guo:2019kdc,Liu:2019tjn}.

The $P_{cs}$ states have attracted less attention. The $P_{cs}(4338)$ and $P_{cs}(4459)$ are associated to $\bar{D}\Xi_c$ and $\bar{D}^*\Xi_c$ molecular states by analogy to the $\bar{D}\Sigma_c$ and $\bar{D}^*\Sigma_c$ structures of the $P_c$ states~\cite{Shen:2025jmy,Ke:2023nra,Azizi:2023iym,Zhu:2022wpi,Chen:2022wkh,Giachino:2022pws,Meng:2022wgl,Xiao:2019gjd,Yan:2022wuz,Wang:2022mxy}. In Ref.~\cite{Ortega:2022uyu}, starting from a quark model structure, the molecular substructure was also studied, associating the $P_{cs}(4459)$ to $\bar{D}\Xi_c$ mostly, and the $P_{cs}(4438)$ to a mixture of many molecular configurations. However, there is a drastic difference of the $P_c$ and $P_{cs}$ states concerning the molecular structure, because while the $P_c$ states are mostly single channel states, with very small admixture with other coupled channels, this is not the case for the $P_{cs}$ states, which was earlier noted in~\cite{Chen:2022wkh}. This issue has been further elaborated in~\cite{Feijoo:2022rxf}. Indeed, by looking at the coupled channels interaction for the $P_c$ states studied in~\cite{Xiao:2019aya}, one observes that the $\bar{D}\Sigma_c$ and $\bar{D}\Lambda_c$ channels do not couple~(see Eq.~(3) of~\cite{Xiao:2019aya}), and similarly $\bar{D}^*\Sigma_c$ and $\bar{D}^*\Lambda_c$ do not mix~(see Eqs.~(4) and~(6) of~\cite{Xiao:2019aya}). On the other hand, the $\bar{D}\Xi_c$ and $\bar{D}_s\Lambda_c$ states couple strongly in the formation of $P_{cs}$ states~(see Table~1 of~\cite{Feijoo:2022rxf}), and so do the $\bar{D}^*\Xi_c$ and $\bar{D}_s^*\Lambda_c$ states~(see Table~2 of~\cite{Feijoo:2022rxf}). The coupled channels are very important in molecular formation, since the presence of a coupled channel implies an additional interaction in the other channel proportional to the square of the transition potential~\cite{Hyodo:2013nka,Aceti:2014ala}. The transition potential in the case of the $\bar{D}_s\Lambda_c$ and $\bar{D}\Xi_c$, or $\bar{D}_s^*\Lambda_c$ and $\bar{D}^*\Xi_c$ channels is very strong and distorts the picture of the $P_{cs}$ states compared with that of the $P_c$ states. As a consequence of that, a picture emerged from the coupled channel approach of~\cite{Feijoo:2022rxf}, in which the $P_{cs}(4338)$ state was associated to the $\bar{D}^*\Xi_c$~(with some mixture of $\bar{D}^*_s\Lambda_c$), and the $P_{cs}(4459)$ to $\bar{D}\Xi_c^\prime$. Two more states emerged from that picture, a state around 4565~MeV coupling practically only to $\bar{D}^*\Xi_c^\prime$ and another one coupling strongly to $\bar{D}\Xi_c$~(and also to $\bar{D}_s\Lambda_c$) which appears around 4200~MeV. The strong effect of the coupled channels lowers the energy of the state coupling to $\bar{D}\Xi_c$ with respect to its analog $P_c$ state that couples to $\bar{D}\Sigma_c$. The coupled channels helping to build the $P_{cs}(4200)$ state are $\eta_c\Lambda$, $\bar{D}_s\Lambda_c$, $\bar{D}\Xi_c$ and $\bar{D}\Xi_c^\prime$, and the couplings of the resonance to these channels are shown in Table~\ref{table:1}, together with the wave functions at the origin $g_iG_i$~\cite{Gamermann:2009uq}.
\begin{table*}\label{table:1}
	
	\caption{Couplings and wave functions at the origin~[in MeV] for the different channels\footnote{$G^{II}_i$ in the Table~\ref{table:1} is the $G$ function evaluated in the second Riemann sheet at the pole of the state. It is defined as $G^{II}_i=G^{I}_i$, for the closed channels, $\bar{D}_s\Lambda_c$, $\bar{D}\Xi_c$ and $\bar{D}\Xi_c^\prime$. $G^{II}_i(\sqrt{s})=G^{I}_i(\sqrt{s})+i\frac{M_B}{2\pi\sqrt{s}}q_{on}$, for the open channel $\eta_c\Lambda$. $q_{on}=\lambda^{1/2}(s,m_m^2,M_B^2)/(2\sqrt{s})$, with $m_m$, $M_B$ the masses of the meson, baryon in the loop, and $G^{I}$ the ordinary meson baryon loop function.}}
	\begin{tabular}{c|lcccc}
		\hline 
		\hline
		Poles           &       ~            & $\eta_c\Lambda$    & $\bar{D}_s\Lambda_c$  & $\bar{D}\Xi_c$   & $\bar{D}\Xi_c^\prime$      \\
		\hline
		{$4198.94+i0.11$} 
		&     ~~$g_i$              & ${0.12-i0.00}$       & $3.01-i0.01$      & $4.85+i0.01$    & $0.01-i0.03$          \\
		$q_{\text{max}}=600$          &    ~~$g_iG^{II}_i$   & ${-0.35+i1.01}$     & $-19.24+i0.05$  & $-20.35-i0.07$ & $-0.03+i0.09$         \\
        \hline
		{$4220-i0.14$} 
		&     ~~$g_i$              & ${0.12+i0.00}$       & $2.69+i0.01$      & $4.57-i0.02$    & $0.00+i0.04$          \\
		$q_{\text{max}}=550$          &    ~~$g_iG^{II}_i$   & ${-0.00-i1.15}$     & $-17.88-i0.06$  & $-17.53+i0.07$ & $-0.00-i0.09$         \\
        \hline
		{$4159.53+i0.06$} 
		&     ~~$g_i$              & ${0.10-i0.01}$       & $3.72-i0.01$      & $5.75-i0.01$    & $0.02-i0.02$          \\
		$q_{\text{max}}=650$          &    ~~$g_iG^{II}_i$   & ${-0.59+i0.66}$     & $-21.51+i0.03$  & $-24.64-i0.04$ & $-0.06+i0.07$         \\
		
		\hline
		\hline
        ~      &       ~            & $J/\psi\Lambda$    & $\bar{D}_s^*\Lambda_c$  & $\bar{D}^*\Xi_c$   & $\bar{D}^*\Xi_c^\prime$      \\
		\hline
		{$4337.98+i0.12$} 
		&     ~~$g_i$              & ${0.11-i0.00}$       & $3.17-i0.01$      & $5.07+i0.01$    & $0.01-i0.04$          \\
		$q_{\text{max}}=600$          &    ~~$g_iG^{II}_i$   & ${-0.13+i1.07}$     & $-18.57+i0.06$  & $-19.94-i0.07$ & $-0.01+i0.10$         \\
        \hline
		{$4360.69-i0.15$} 
		&     ~~$g_i$              & ${0.12+i0.00}$       & $2.82+i0.01$      & $4.80-i0.02$    & $-0.01+i0.04$          \\
		$q_{\text{max}}=550$          &    ~~$g_iG^{II}_i$   & ${0.41-i1.21}$     & $-17.32-i0.06$  & $-17.17+i0.07$ & $0.03-i0.10$         \\
        \hline
		{$4297.97+i0.07$} 
		&     ~~$g_i$              & ${0.10+i0.00}$       & $3.91+i0.01$      & $6.02-i0.01$    & $0.02+i0.03$          \\
		$q_{\text{max}}=650$          &    ~~$g_iG^{II}_i$   & ${-0.45-i0.76}$     & $-20.83-i0.04$  & $-24.01+i0.05$ & $-0.05-i0.08$         \\
		
		\hline
		\hline
	\end{tabular}
\end{table*}
As we can see, the main building blocks are $\bar{D}\Xi_c$ and $\bar{D}_s\Lambda_c$. The $\eta_c\Lambda$ couples weakly to the resonance, but having a mass lower than the resonance mass~(4099~MeV), it is the natural decay channel of that state, but the $\bar{D}_s\Lambda_c$ is the channel which we will consider in order to produce the resonance. Since the coupling to $\eta_c\Lambda$ is so small, the width of the resonance is also small, around 200~keV.

One might think that the small coupling to $\eta_c\Lambda$, the channel where the resonance is suggested to be observed, should make the signal of this resonance very small in any reaction, but we shall see that this is not the case here, because in the reaction that we propose we produce the resonance with the $\bar{D}_s\Lambda_c$ channel off shell, and hence, the production of the resonance is not linked to the $\eta_c\Lambda$ channel. Once produced it will decay, in this case into $\eta_c\Lambda$, but the probability to find the resonance in $\eta_c\Lambda$ is the production probability of the $P_{cs}$, because this is the only decay channel. It is then not surprising that we obtain a branching ratio for the $P_{cs}(4200)$ state which is relatively large.

We propose to search for this resonance in $\Lambda_b\to\phi\eta_c\Lambda$ decay. The reason to propose this reaction is that the $\Lambda_b^0\to\phi D_s^-\Lambda_c^+$ decay has already been observed in~\cite{LHCb:2025lwm}. The reaction is $\Lambda_b^0\to\Lambda_c^+D_s^-K^+K^-$, but the $K^+K^-$ come mostly from $\phi$ decay. Experimentally one finds~\cite{LHCb:2025lwm}
\begin{equation}
	\dfrac{\mathcal{B}r(\Lambda_b\to\Lambda_c\bar{D}_s\phi)}{\mathcal{B}r(\Lambda_b\to\Lambda_c\bar{D}_s)}\sim 10^{-2},
\end{equation}
which together with the branching fraction of the $\Lambda_b\to\Lambda_c\bar{D}_s$, which is of the order of $10^{-2}$~\cite{ParticleDataGroup:2024cfk}, gives
\begin{equation}\label{Eq:Br}
	\mathcal{B}r(\Lambda_b\to\Lambda_c\bar{D}_s\phi)\sim 10^{-4}.
\end{equation}

\section{Formalism}\label{sec2}

\begin{figure}[htbp]
	\centering
	
	\includegraphics[scale=0.65]{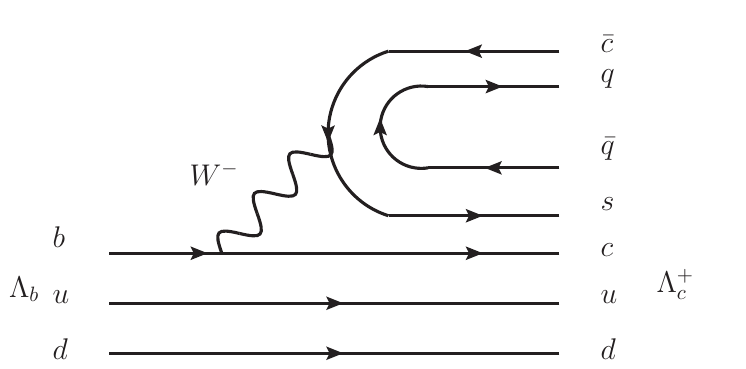}
	
	\caption{$\Lambda_b\to\Lambda_c\bar{c}s$, together with hadronization of the $s\bar{c}$ component.}\label{fig:LbtoLccqqbars}
\end{figure}

It is interesting to see how the $\Lambda_b^0\to\Lambda_c^+D_s^-\phi$ reaction can proceed at the microscopic level. We can see that in Fig.~\ref{fig:LbtoLccqqbars} with external emission. The $s\bar{c}$ pair in Fig.~\ref{fig:LbtoLccqqbars} hadronizes as
\begin{equation}
	s\bar{c}\Rightarrow\sum_is\bar{q}_iq_i\bar{c}=\sum_i M_{3i}M^\prime_{i4}=(MM^\prime)_{34},\nonumber
\end{equation}
where $M_{ij}$, $M^\prime_{ij}$ are $q_i\bar{q}_j$ matrices written in terms of mesons, pseudoscalar or vectors. We have

\begin{equation}\label{Eq:P}
	P=\left(\begin{array}{cccc}
		\frac{\pi^0}{\sqrt{2}}+\frac{\eta}{\sqrt{3}}+\frac{\eta'}{\sqrt{6}}& \pi^{+}&K^{+}&\bar{D}^0\\
		\pi^{-}&-\frac{\pi^0}{\sqrt{2}}+\frac{\eta}{\sqrt{3}}+\frac{\eta'}{\sqrt{6}}&K^0&D^{-} \\
		K^{-} & \bar{K}^0 & -\frac{\eta}{\sqrt{3}} +\frac{2\eta'}{\sqrt{6}} & D_s^{-} \\
		D^0 & D^{+} & D_s^{+} & \eta_c
	\end{array}\right)
\end{equation}
\begin{equation}\label{Eq:V}
	V=\left(\begin{array}{cccc}
		\frac{\rho^0}{\sqrt{2}}+\frac{\omega}{\sqrt{2}} ~~~& \rho^{+} & K^{*+} & \bar{D}^{* 0} \\
		\rho^{*-} & -\frac{\rho^0}{\sqrt{2}}+\frac{\omega}{\sqrt{2}} ~~~& K^{* 0} ~~~& D^{*-}~~~ \\
		K^{*-} & \bar{K}^{* 0} & \phi & D_s^{*-} \\
		D^{* 0} & D^{*+} & D_s^{*+} & J / \psi
	\end{array}\right) .
\end{equation}
We need to hadronize $s\bar{c}$ into a pseudoscalar meson and a vector and we can have
\begin{align}
	(PV)_{34}:  ~~~ s\bar{c}\rightarrow(PV)_{34},\\
	(VP)_{34}:  ~~~ s\bar{c}\rightarrow(VP)_{34},
\end{align}
\begin{align}\label{Eq:PV}
(PV)_{34}=(K^-\bar{D}^{*0}+\bar{K}^0\bar{D}^{*-}-\frac{1}{\sqrt{3}}\eta D_s^{*-}+\cdot\cdot\cdot),
\end{align}
\begin{align}\label{Eq:VP}
(VP)_{34}=(K^{*-}\bar{D}^{0}+\bar{K}^{*0}D^{-}+\phi D_s^{-}+\cdot\cdot\cdot),
\end{align}
where the dots indicate states with extra $c\bar{c}$ quarks not relevant in our study. We can see that the $(VP)_{34}$ contribution gives rise to $\Lambda_b^0\to\Lambda_c^+\phi D_s^-$.

\begin{figure}[htbp]
	\centering
	
	\includegraphics[scale=0.65]{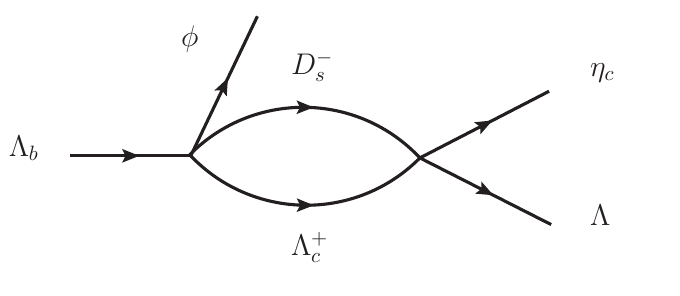}
	
	\caption{Rescattering of $D_s^-\Lambda_c^+$ to $\eta_c\Lambda$, mediated by the resonance $P_{cs}(4200)$.}\label{fig:rescattering}
\end{figure}

Once $\Lambda_c^+D_s^-$ is produced, we can obtain $\Lambda_b^0\to\phi\eta_c\Lambda$ through rescattering of the $\Lambda_c^+D_s^-$ component as shown in Fig.~\ref{fig:rescattering}.

\begin{figure}
	\subfigure[]{
		\centering
			\includegraphics[scale=0.65]{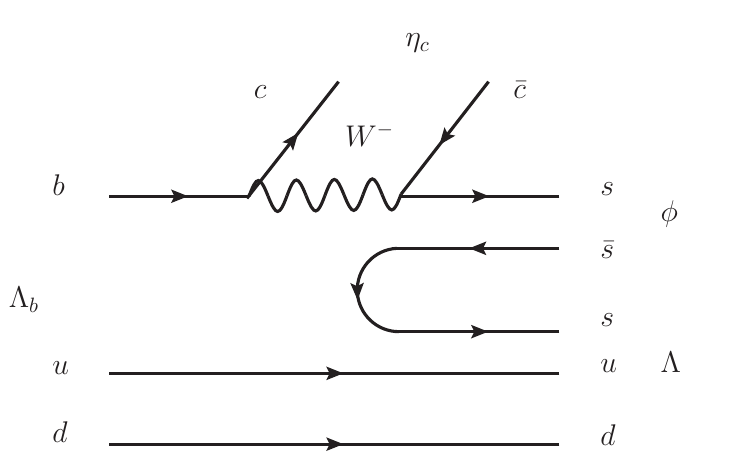}
		}
	\subfigure[]{
		\centering
			\includegraphics[scale=0.65]{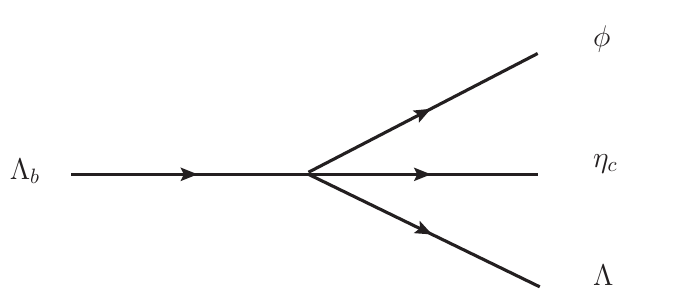}
		}
	\caption{(a) Mechanism to produce $\eta_c\phi\Lambda$ in $\Lambda_b$ decay through internal emission. (b) The mechanism depicted as a tree level diagram.}\label{fig:internal_emission}
\end{figure}

There is another mechanism to produce $\eta_c\Lambda\phi$ through internal emission, as shown in Fig.~\ref{fig:internal_emission}.
The mechanism of Fig.~\ref{fig:internal_emission} is color suppressed and would have a reduction factor of about $1/N_c$ with respect to $\Lambda_b^0\to\phi D_s^-\Lambda_c^+$. We, thus, have the tree level production of Fig.~\ref{fig:internal_emission}(b) together with the rescattering mechanism of Fig.~\ref{fig:rescattering}, corresponding to a production amplitude for $\Lambda_b\to\phi\eta_c\Lambda$ 
\begin{equation}\label{Eq:t}
t=A\beta+AG_{\bar{D}_s\Lambda_c}(M_{\text{inv}}(\eta_c\Lambda))t_{\bar{D}_s\Lambda_c,\eta_c\Lambda}(M_{\text{inv}}(\eta_c\Lambda)),
\end{equation}
with $\beta\sim 1/N_c$ and $A$ the vertex $\Lambda_b^0\to\phi D_s^-\Lambda_c^+$. In Eq.~(\ref{Eq:t}) $G$ is the loop function of the $D_s^-\Lambda^+_c$ propagation of Fig.~\ref{fig:rescattering}, which is calculated in~\cite{Feijoo:2022rxf} and is regularized with cut off regularization. We use the information of~\cite{Feijoo:2022rxf}, but the needed information is already available in Table~\ref{table:1}. Indeed, we can write the $t_{\bar{D}_s\Lambda_c,\eta_c\Lambda}$ amplitude corresponding to the $P_{cs}(4200)$ resonance as
\begin{equation}\label{Eq:scattering_t}
	t_{\bar{D}_s\Lambda_c,\eta_c\Lambda}=\dfrac{g_{\bar{D}_s\Lambda_c}g_{\eta_c\Lambda}}{M_{\text{inv}}(\eta_c\Lambda)-M_{P_{cs}}+i\frac{\Gamma_{P_{cs}}}{2}}.
\end{equation}
Then in the contribution $G_{\bar{D}_s\Lambda_c}t_{\bar{D}_s\Lambda_c,\eta_c\Lambda}$ we will have $G_{\bar{D}_s\Lambda_c}g_{\bar{D}_s\Lambda_c}$, which is the $\bar{D}_s\Lambda_c$ wave function at the origin of the $\bar{D}_s\Lambda_c$ channel and is tabulated in Table~\ref{table:1}, together with $g_{\eta_c\Lambda}$ and $\Gamma_{P_{cs}}/2$ (imaginary part of the pole position, 0.11~MeV). We need the value of A in Eq.~(\ref{Eq:t}) to obtain the branching ratio in absolute terms. We evaluate it as follows.

\begin{figure}[htbp]
	\centering
	
	\includegraphics[scale=0.65]{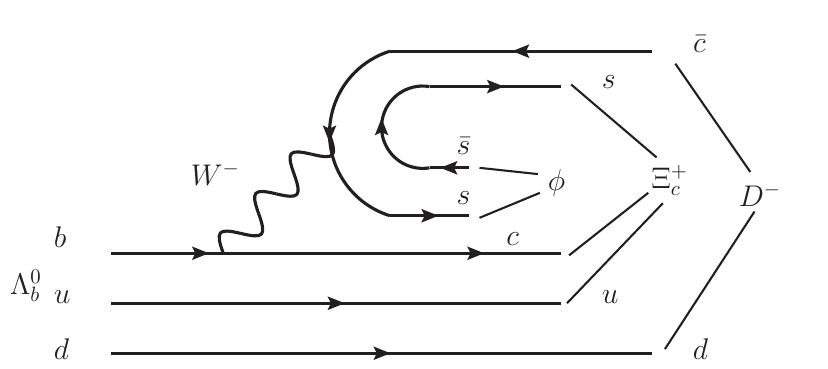}
	
	\caption{Possible mechanism for $\Lambda_b^0\to\Xi_c^+\phi D^-$.}\label{fig:new_mechanism}
\end{figure}

It is interesting to mention here that, while the $\bar{D}\Xi_c$ channel is the most important component, together with $\bar{D}_s\Lambda_c$, as shown in Table.\ref{table:1}, that channel is not produced in the weak decay, as one can see in Eqs.~(\ref{Eq:PV}) and~(\ref{Eq:VP}). It is also easy to see that in Figs.~\ref{fig:LbtoLccqqbars} and~\ref{fig:internal_emission}(a) because the $ud$ quarks of $\Lambda_b$, produced in $I=0$, act as spectators in the reaction and $\Xi_c$ has $csq$, one extra $s$ quark. One can think of some different topologies as the one shown in Fig.~\ref{fig:new_mechanism}. There the $d$ quark of the $\Lambda_b$ is directly involved in the weak decay since it is forced to transfer to the $D^-$ meson with the $\bar{c}$ quark produced from the $W^-$. Similarly, one $s$ quark from the hadronization of the $\bar{c}s$ component produced by the $W^-$, becomes one quark of the $\Xi_c^+$. Such mechanisms, although possible, involve large momentum transfers in the evaluation of the transition matrix element, involving form factors, with the consequent reduction of their strength, and do not compete with the straightforward one where the three quarks of the $\Lambda_b$ baryon remain in the final baryon, and two of them act as spectators~\cite{Miyahara:2015cja,Miyahara:2016yyh}.

\subsection{Extraction of A from the $\Lambda_b^0\to\phi D_s^-\Lambda_c^+$ reaction}
We must contract the polarization of the $\phi$, $\vec{\varepsilon}_\phi$, with some vector. We choose one combination $\vec{\varepsilon}_\phi\cdot\vec{\sigma}$ with $\vec{\sigma}$ the spin operator acting on the baryons. Other choices can be done, but since this is propagated in the diagrams of Fig.~\ref{fig:rescattering}, it simply leads to a factor which is common to all processes and disappears in the ratios that we construct.

The decay width for the process $\Lambda_b^0\to\phi D_s^-\Lambda_c^+$ is given by
\begin{equation}\label{Eq:dwidh_L2phiDsLc}
	\frac{d\Gamma}{dM_{\text{inv}}(\bar{D}_s\Lambda_c)}=\frac{1}{(2\pi)^3}\frac{2M_{\Lambda_c}2M_{\Lambda_b}}{4M_{\Lambda_b}^2}p_\phi\tilde{p}_{\bar{D}_s}\overline{\sum}\sum|t^\prime|^2,
\end{equation}
with $t^\prime=A\vec{\sigma}\cdot\vec{\epsilon}_\phi$, and
\begin{equation}
	p_\phi=\dfrac{\lambda^{1/2}(M_{\Lambda_b}^2,m_\phi^2,M^2_{\text{inv}}(\bar{D}_s\Lambda_c))}{2M_{\Lambda_b}},
\end{equation}
\begin{equation}
	\tilde{p}_{\bar{D}_s}=\dfrac{\lambda^{1/2}(M^2_{\text{inv}}(\bar{D}_s\Lambda_c),m_{\bar{D}_s}^2,M^2_{\Lambda_c})}{2M_{\text{inv}}(\bar{D}_s\Lambda_c)}.
\end{equation}
The sum and average over the spins of the $\Lambda_b$, $\Lambda_c$ gives
\begin{align}
\overline{\sum}\sum|t^\prime|^2=&\frac{1}{2}\sum_{\epsilon_{\text{pol}}}\sum_m\sum_{m^\prime}\left<m|\sigma_i\epsilon_i|m^\prime\right>\left<m^\prime|\sigma_j\epsilon_j|m\right>A^2 \nonumber\\
=&\frac{1}{2}\sum_{\epsilon_{\text{pol}}}\sum_m\left<m|\delta_{ij}+i\epsilon_{ijk}\sigma_k|m\right>A^2\epsilon_i\epsilon_j \nonumber\\
=&A^2\delta_{ij}\delta_{ij}\nonumber\\
=&3A^2. 
\end{align}
Integrating Eq.~(\ref{Eq:dwidh_L2phiDsLc}) over $M_{\text{inv}}(\bar{D}_s\Lambda_c)$ and comparing the result with the branching fraction of Eq.~(\ref{Eq:Br}) we obtain the value of $A^2/\Gamma_{\Lambda_b}$ which is
\begin{equation}\label{Eq:A2overGamma}
	\frac{A^2}{\Gamma_{\Lambda_b}}=2.23\times10^{-10}~\text{MeV}^{-3}.
\end{equation}

\subsection{Branching ratio for $\Lambda_b\to\phi\eta_c\Lambda$}
Next, we can evaluate the branching ratio for $\Lambda_b\to\phi\eta_c\Lambda$. We have
\begin{equation}\label{Eq:dwidh_L2phietacL}
	\frac{d\Gamma}{dM_{\text{inv}}(\eta_c\Lambda)}=\frac{1}{(2\pi)^3}\frac{M_{\Lambda}}{M_{\Lambda_b}}p_\phi\tilde{p}_{\eta_c}|t|^2\cdot3,
\end{equation}
with $t$ given by Eq.~(\ref{Eq:t}), with
\begin{equation}
	p_\phi=\dfrac{\lambda^{1/2}(M_{\Lambda_b}^2,m_\phi^2,M^2_{\text{inv}}(\eta_c\Lambda))}{2M_{\Lambda_b}},
\end{equation}
\begin{equation}
	\tilde{p}_{\eta_c}=\dfrac{\lambda^{1/2}(M^2_{\text{inv}}(\eta_c\Lambda),m_{\eta_c}^2,M^2_{\Lambda})}{2M_{\text{inv}}(\eta_c\Lambda)},
\end{equation}
and to calculate the branching ratio we divide by $\Gamma_{\Lambda_b}$, and for that we use $A^2/\Gamma_{\Lambda_b}$ of Eq.~(\ref{Eq:A2overGamma}).

\section{Results}\label{sec3}

\begin{figure}[htbp]
	\centering
	
	\includegraphics[scale=0.65]{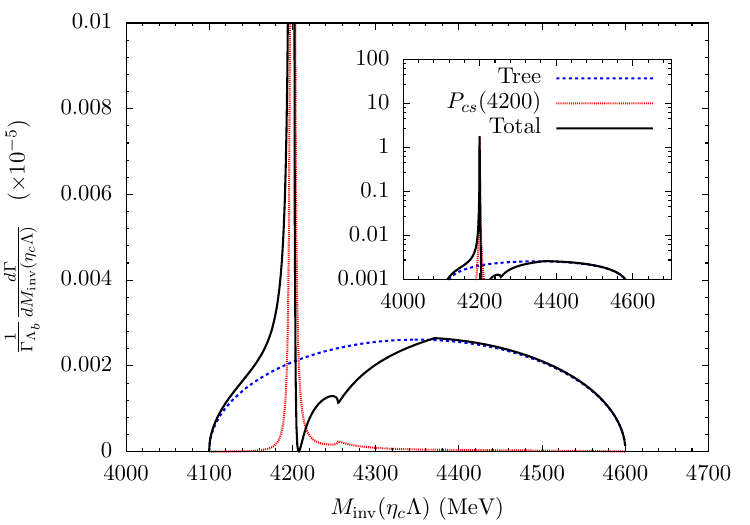}
	
	\caption{$\frac{1}{\Gamma_{\Lambda_b}}\frac{d\Gamma}{dM_{\text{inv}}(\eta_c\Lambda)}$ as a function of $M_{\text{inv}}(\eta_c\Lambda)$ for the $\Lambda_b\to\phi\eta_c\Lambda$.}\label{fig:width}
\end{figure}

\begin{figure}[htbp]
	\centering
	
	\includegraphics[scale=0.65]{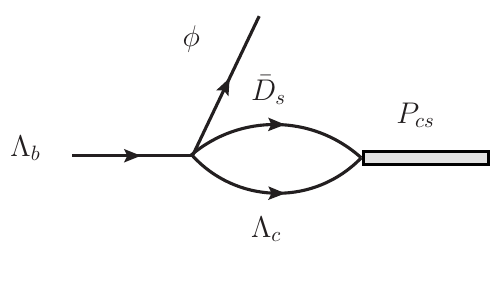}
	
	\caption{Coalescence $P_{cs}$ production in $\Lambda_b\to\phi P_{cs}$ driven by the $\bar{D}_s\Lambda_c$ intermediate state.}\label{fig:Pcs_produce}
\end{figure}

In Fig.~\ref{fig:width}~~\footnote{Around 4350~MeV, far away from the resonance peak, we take the prescription of~\cite{Roca:2025zyi} to write $G_{\bar{D}_s\Lambda_c}=\text{Re}(G_{\bar{D}_s\Lambda_c})+i \text{Im}(G_{\bar{D}_s\Lambda_c})$ with $q_{\text{max}}=600$~MeV from~\cite{Feijoo:2022rxf}, taking $\text{Re}(G_{\bar{D}_s\Lambda_c})=0$ if it becomes positive and the exact expression for $\text{Im}(G_{\bar{D}_s\Lambda_c})$. Alternatively, we can use $gG$ for $\bar{D}_s\Lambda_c$ at the pole, tabulated in Table~\ref{table:1}, and the results are basically the same.}, we observe a very narrow peak corresponding to the narrow $P_{cs}(4200)$ resonance. There is also a small cusp appearing at the opening of the $\bar{D}_s\Lambda_c$ channel around 4250~MeV. If we integrate over the whole $\eta_c\Lambda$ invariant mass range we obtain
\begin{equation}
	\dfrac{\Gamma_{\Lambda_b\to\phi\eta_c\Lambda}}{\Gamma_{\Lambda_b}}\simeq3.8\times10^{-5},
\end{equation}
well within measurable rates for $\Lambda_b$ decays reported in the PDG~\cite{ParticleDataGroup:2024cfk}. It is instructive to see the rate obtained for the resonance peak. Integrating in the range $M_{P_{cs}}-2\Gamma_{P_{cs}}$, $M_{P_{cs}}+2\Gamma_{P_{cs}}$, we obtain
\begin{equation}\label{Eq:Br_Pcs_1}
	\mathcal{B}r[\Lambda_b\to\phi P_{cs}(4200)]\simeq2.46\times 10^{-5}.
\end{equation}
It is interesting to see that this rate does not depend on $g_{\eta_c\Lambda}$, the small coupling of the resonance to the $\eta_c\Lambda$ decay channel. Indeed, the resonance part of the width will come from the $Gt$ term of Eq.~(\ref{Eq:t}). Then
\begin{align}
	|t_{\bar{D}_s\Lambda_c,\eta_c\Lambda}|^2=&\dfrac{g^2_{\bar{D}_s\Lambda_c}g^2_{\eta_c\Lambda}A^2G^2_{\bar{D}_s\Lambda_c}}{(M_{\text{inv}}(\eta_c\Lambda)-M_{P_{cs}})^2+\left(\frac{\Gamma_{P_{cs}}}{2}\right)^2} \nonumber\\
	=&A^2(g_{\bar{D}_s\Lambda_c}G_{\bar{D}_s\Lambda_c})^2g^2_{\eta_c\Lambda}\left(-\frac{2}{\Gamma_{P_{cs}}}\right) \nonumber\\
	&\times \text{Im}\left(\frac{1}{M_{\text{inv}}(\eta_c\Lambda)-M_{P_{cs}}+i\frac{\Gamma_{P_{cs}}}{2}}\right)\nonumber\\
	\simeq&A^2(g_{\bar{D}_s\Lambda_c}G_{\bar{D}_s\Lambda_c})^2g^2_{\eta_c\Lambda}\left(-\frac{2}{\Gamma_{P_{cs}}}\right)(-\pi) \nonumber\\
	&\times\delta(M_{\text{inv}}(\eta_c\Lambda)-M_{P_{cs}}),
\end{align}
where we have used that $\text{Im}(x-x_0+i\epsilon)=-\pi\delta(x-x_0)$, which upon integration over $M_{\text{inv}}$, and summing over spins and $\phi$ polarizations, gives
\begin{align}\label{Eq:Br_method2}
	\mathcal{B}r[\Lambda_b\to\phi P_{cs}(4200)]=\frac{1}{(2\pi)^3}\frac{M_\Lambda}{M_{\Lambda_b}}p_\phi\tilde{p}_{\eta_c}3A^2\nonumber\\
	\times(g_{\bar{D}_s\Lambda_c}G_{\bar{D}_s\Lambda_c})^2g^2_{\eta_c\Lambda}\frac{2\pi}{\Gamma_{P_{cs}}},
\end{align}
with $p_\phi$, $\tilde{p}_{\eta_c}$ calculated at $M_{\text{inv}}(\eta_c\Lambda)=M_{P_{cs}}$. Since $\Gamma_{P_{cs}}$ is proportional to $g_{\eta_c\Lambda}^2$, the coupling $g_{\eta_c\Lambda}$ disappears in the formula of Eq.~(\ref{Eq:Br_method2}). One can go one step forward and considering that 
\begin{equation}
	\Gamma_{P_{cs}}=\frac{1}{8\pi}\frac{2M_{\Lambda}2M_{P_{cs}}}{M^2_{P_{cs}}}g^2_{\eta_c\Lambda}\tilde{p}_{\eta_c},
\end{equation}
we obtain
\begin{equation}\label{Eq:Br_Pcs_2}
	\Gamma_{\Lambda_b\to\phi P_{cs}}=\frac{1}{8\pi}\frac{2M_{\Lambda_b}2M_{P_{cs}}}{M^2_{\Lambda_b}}3A^2(g_{\bar{D}_s\Lambda_c}G_{\bar{D}_s\Lambda_c})^2p_{\phi},
\end{equation}
which corresponds to the production of the $P_{cs}$ resonance with coalescence production depicted in Fig.~\ref{fig:Pcs_produce}.
Hence, it is proved that the small coupling of the resonance $P_{cs}(4200)$ to $\eta_c\Lambda$ is no obstacle to having a large branching ratio. Only the small width of $\Gamma_{P_{cs}}=0.22$~MeV will pose experimental difficulties, but small widths of this type are measured in experiments of LHCb like in the observation of the $T_{cc}$ states~\cite{LHCb:2021vvq,LHCb:2021auc}.

The result obtained with Eq.~(\ref{Eq:Br_Pcs_2}) gives $\mathcal{B}r[\Lambda_b\to\phi P_{cs}]=2.47\times10^{-5}$, basically the same one calculated in Eq.~(\ref{Eq:Br_Pcs_1}). If the resolution of an experiment is bigger than this width, the spectrum will be seen wider but the integrated width will be the same. Branching ratios of $10^{-5}$ are common in $\Lambda_b$ decay~\cite{ParticleDataGroup:2024cfk}, where even branching ratios of $10^{-7}$ are reported. Hence, the proposed experiment is well within the LHCb capabilities at present, and certainly in future runs of the LHCb.

\subsection{Determination of uncertainties}\label{sec3A}

We discuss here possible uncertainties in the results. The position of the pole, the couplings, $G$ functions, and as a consequence the production rate of the reaction, depend on the cutoff $q_{\text{max}}$ used to regularize the $G$ functions in the approach. We vary this value now to estimate uncertainties in the results. The value of $q_{\text{max}}$ was tuned to get the two states, $P_{cs}(4338)$, $P_{cs}(4459)$ with the right mass in Ref.~\cite{Feijoo:2022rxf}, the $P_{cs}(4338)$ corresponding mostly to $\bar{D}^*\Xi_c$ and the $P_{cs}(4459)$ to $\bar{D}\Xi_c^\prime$. The choice was made to obtain the mass of the $P_{cs}(4338)$, while the $P_{cs}(4459)$ state appeared at 4423~MeV, about 35~MeV below the nominal mass. This means that we must admit easily about 30~MeV uncertainty in the mass of the $P_{cs}(4200)$ state that we obtain. We have repeated the calculations of Ref.~\cite{Feijoo:2022rxf} for the $PB~(1/2^+)$ and $VB~(1/2^-)$ sectors and have seen that a change in the mass of the $P_{cs}(4200)$ and $P_{cs}(4338)$ of 30~MeV is obtained using $q_{\text{max}}=550$~MeV and 650~MeV. We show the results in Table~\ref{table:1}, together with the couplings and wave function at the origin. As one can see, the changes in the mass of the $P_{cs}(4200)$ are about 30~MeV, but the changes in the couplings are moderate. We have then reevaluated the mass distributions and integrated rates. The mass distributions are similar as that in Fig.~\ref{fig:width}, but with the peak displaced to the new value of the $P_{cs}$ mass. As to the production rate, we show the new results in Table~\ref{table:2}, and see that the changes with respect to the central value are of the order of 20\%. 

\begin{table}
	\caption{Pole position and branching ratio with different values of $q_{\text{max}}$ [in units of MeV]}\label{table:2}
	\begin{tabular}{c|ccc}
		\hline 
		\hline
		~           &       $q_{\text{max}}=550$            & $q_{\text{max}}=600$   & $q_{\text{max}}=650$   \\
		\hline
		Pole position 
		&     ${4220-i0.14}$              & ${4198.94+i0.11}$       & ${4160+i0.66}$          \\
		$\mathcal{B}r[\Lambda_b\to\phi P_{cs}]$          &    ${2.21\times10^{-5}}$   & ${2.47\times10^{-5}}$     & ${3.16\times10^{-5}}$  \\
		
		\hline
		\hline
	\end{tabular}
\end{table}

\section{ Conclusions }
We propose here a reaction to detect a $P_{cs}$ state predicted around $4200$~MeV, not yet observed. This $P_{cs}$ state at lower energy than expected from a comparison of the $P_c$ states with the $P_{cs}$ ones, stems because of the role played by coupled channels in the $P_{cs}$ case which do not appear in the $P_c$ case. The reaction proposed is $\Lambda_b\to \phi \eta_c \Lambda$. The motivation to propose this reaction is that the $\Lambda_b^0\to\phi D_s^- \Lambda_c^+$ reaction has already been observed and the predicted $P_{cs}(4200)$ state couples strongly to $D_s^- \Lambda_c$ and $\bar{D} \Xi_c$. The coupling of the $P_{cs}(4200)$ state to $\eta_c \Lambda$ is very small, and this being the only decay channel, makes the width of that state very small. However, this is no obstacle to having large production rates, because the production of the resonance is driven primarily by the $\Lambda_b^0\to\phi D_s^-  \Lambda_c^+$ reaction, where the resonance is produced which later decays into the $\eta_c \Lambda$ channel. We obtain branching ratios for the production of this resonance of the order of $10^{-5}$, well within present LHCb capabilities, and we anticipate that it could be easily studied in new runs of the LHCb. This rate is very stable when evaluating uncertainties in the mass of the $P_{cs}(4200)$ state. The observation of this $P_{cs}$ state with a mass lower than expected from comparison to the $P_c$ states, would shed valuable light on the nature of the $P_c$ and $P_{cs}$ resonances and the important role played by coupled channels in the structure of hadrons and hadronic reactions.
	
\section*{Acknowledgments}

This work was supported by the National Key R\&D Program of China (Grant No. 2024YFE0105200), the Natural Science Foundation of Henan (Grant No. 252300423951), the National Natural Science Foundation of China (Grant No. 12475086), and the Zhengzhou University Young Student Basic Research Projects for PhD students (Grant No. ZDBJ202522). Wen-Tao Lyu acknowledges the support of the China Scholarship Council. This work is also partly supported by the Spanish Ministerio de Economia y Competitividad (MINECO) and European FEDER funds under Contracts No. FIS2017-84038-C2-1-PB, PID2020-112777GB-I00, and by Generalitat Valenciana under contract PROMETEO/2020/023. This project has received funding from the European Union Horizon 2020 research and innovation program under the program H2020-INFRAIA-2018-1, grant agreement No. 824093 of the STRONG-2020 project.

\end{document}